\documentclass[twocolumn,jcp,superscriptaddress]{revtex4}

\usepackage{graphicx}

\renewcommand{\epsilon}{\varepsilon}
\newcommand{\figurewidth}{0.35\textwidth}
\newcommand{\narrowfigurewidth}{0.23\textwidth}

\begin{document}
\title{Langevin Dynamics Simulations of Polymer Translocation through Nanopores}
\author{Ilkka Huopaniemi}
\affiliation{Laboratory of Physics, Helsinki University of Technology, P.O. Box 1100, FIN-02015 
HUT, Espoo, Finland}
\author{Kaifu Luo }
\email{luokaifu@yahoo.com}
\affiliation{Laboratory of Physics, Helsinki University of Technology, P.O. Box 1100, FIN-02015 
HUT, Espoo, Finland}
\author{Tapio Ala-Nissila}
\affiliation{Laboratory of Physics, Helsinki University of Technology, P.O. Box 1100, FIN-02015 
HUT, Espoo, Finland}
\author{See-Chen Ying}
\affiliation{Department of Physics, Box 1843, Brown University, Providence, RI 
02912-1843, U.S.A.}

\date{\today}

\begin{abstract}
We investigate the dynamics of polymer translocation 
through a nanopore using two-dimensional Langevin dynamics simulations. 
In the absence of external driving force, we consider a polymer which is initially placed
in the middle of the pore and study the escape time $\tau_e$ required for the polymer 
to completely exit the pore on either side. The distribution of the escape times is wide 
and has a long tail. We find that $\tau_e$ scales with the chain 
length $N$ as $\tau_e \sim N^{1+2\nu}$, where $\nu $ is the Flory exponent.
For driven translocation, we concentrate 
on the influence of the friction
coefficient $\xi$, the driving force $E$ and the length of the chain $N$
on the translocation time $\tau$, which is defind as the time duration between the first 
monomer entering the pore and the last monomer leaving the pore. 
For strong driving forces, the distribution of translocation 
times is symmetric and narrow without a long tail and $\tau \sim E^{-1}$. 
The influence of $\xi$ depends on the ratio between the driving and frictional forces. 
For intermediate $\xi$, we find a crossover scaling for $\tau$ with $N$ from $\tau \sim 
N^{2\nu }$ for relatively short chains to $\tau \sim N^{1 + \nu }$ for 
longer chains. 
However, for higher $\xi$, only $\tau \sim N^{1 + \nu }$ is observed 
even for short chains, and there is no crossover behavior. 
This result can be explained by the fact that increasing $\xi$ increases the Rouse 
relaxation time of the chain, in which case even relatively short chains have no 
time to relax during translocation. 
Our results are in good agreement with previous simulations based on the fluctuating 
bond lattice model of polymers at intermediate friction  values, but reveals additional 
features of dependency on friction.
\end{abstract}


\maketitle
\section{Introduction}

The transport of a polymer through a nanopore is associated with an energy 
barrier arising from loss of configurational entropy due to the geometric 
constriction. Such processes are commonly observed in biology, such as DNA 
and RNA translocation across nuclear pores, protein transport through 
membrane channels, and virus injection ~\cite{Alberts,Darnell,Miller}. Moreover, 
translocation processes have various potential technological applications, 
such as rapid DNA sequencing ~\cite{Han,Turner}, gene therapy and controlled 
drug delivery, \textit{etc} ~\cite{Chang}. In addition, the translocation 
dynamics is also a challenging topic in polymer physics. Particularly, the 
scaling of translocation time $\tau $ with the chain length $N$ is an important
 measure of the underlying dynamics of polymer translocation. As a result, 
recently a considerable number of experimental ~\cite{Kasianowicz,Aktson,
Meller2,Henrickson,Meller,Sauer,Meller3,Li,Gershow,Chen,Storm}, 
theoretical~\cite{Storm,Simon,Sung,Park,diMarzio,Muthukumar,Muthu2,Lubensky,
Slonkina,Ambj,Metzler,Ambj2,Ambj3,Gerland,Baumg,Chuang,Kantor,Milchev,Luo1,Luo2} 
and numerical studies ~\cite{Chuang,Kantor,Milchev,Luo1,Luo2,Chern,Loebl,Randel,
Lansac,Kong,Farkas,Tian,Zandi,Guo} have been carried out.

In order to overcome a large entropic barrier in polymer translocation and 
to speed up the translocation, an external driving force is needed in 
experiments, such as an electric field, chemical potential difference, or 
selective adsorption on one side of the membrane. For example, in 1996, 
Kasianowicz \textit{et al}.~\cite{Kasianowicz} reported an elegant experiment 
where an electric field can drive single-stranded DNA and RNA molecules through 
the $\alpha $-hemolysin channel of inside diameter 2 nm and that the passage 
of each molecule is signaled by the blockade in the channel current. In fact, the 
translocation process includes two essential steps. First, one end of the 
polymer enters the pore directed by diffusion and by the action of an 
electric field near the pore. The experimental results show that the ability 
of the polymer to enter the nanopore depends linearly on polymer 
concentration and exponentially on the applied voltage ~\cite{Kasianowicz,Meller}. 
Second, the polymer is translocated from one side of the membrane to the other, 
driven by the electric field. In the experiment for $\alpha$-hemolysin, 
the linear behavior of $\tau $ with the $N$ is observed ~\cite{Kasianowicz,Meller}. 
In addition, an inverse linear 
dependence and an inverse quadratic dependence of the translocation time on 
applied voltage are observed for different experiments ~\cite{Kasianowicz,Meller}.

To overcome the limited voltage range that can be applied across a 
biological pore and the difficulty in analyzing the current variations 
because the shot noise is comparable to the expected signal, recently 
Li \textit{et al.}~\cite{Li,Gershow} showed that a solid-state nanopore could 
also be used for similar experiments. Most Recently, Storm \textit{et al.}~\cite{Storm} 
carried out a set of experiments on double-stranded DNA molecules with various 
lengths that translocate through a solid-state nanopore ~\cite{Chen}. Surprisingly, 
a power-law scaling of the most probable translocation time with the polymer length, 
with an exponent 1.27 was observed, in contrast to the linear behavior observed for 
all experiments on $\alpha $-hemolysin channel.

Inspired by the experiments ~\cite{Kasianowicz,Meller,Storm}, a number of recent 
theories ~\cite{Storm,Simon,Sung,Park,diMarzio,Muthukumar,Muthu2,Lubensky,Slonkina,
Ambj,Metzler,Ambj2,Ambj3,Gerland,Baumg,Chuang,Kantor,Milchev,Luo1,Luo2} 
have been developed for the dynamics of polymer translocation. Even for the field
free case, polymer translocation remains a challenging problem. To 
this end, Sung and Park~\cite{Sung} and Muthukumar~\cite{Muthukumar} considered 
equilibrium entropy of the polymer as a function of the position of the polymer 
through the nanopore. Standard Kramer analysis of diffusion through this entropic 
barrier yields a scaling prediction of the translocation time $\tau \sim 
N^2$ for long chains. However, as Chuang \textit{et al}.~\cite{Chuang} noted, 
this quadratic scaling behavior cannot be correct for a self-avoiding polymer. 
The reason is that the translocation time is shorter than the equilibration time 
of a self-avoiding polymer, $\tau _{equil} \sim N^{1 + 2\nu }$, where $\nu $ is 
the Flory exponent ~\cite{de Gennes,Doi}, thus rendering the concept of equilibrium 
entropy and the ensuing entropic barrier inappropriate for the study of 
translocation dynamics. Chuang\textit{ et al}.~\cite{Chuang} performed numerical 
simulations with Rouse dynamics for a two-dimensional (2D) lattice model to study 
the translocation for both phantom and self-avoiding polymers. They decoupled 
the translocation dynamics from the diffusion dynamics outside the pore by 
imposing the artificial restriction that the first monomer, which is initially 
placed in the pore, is never allowed to cross back out of the pore. Their results 
show that for large $N$, translocation time scales approximately in the same manner 
as equilibration time.

For forced translocation, Sung and Park~\cite{Sung} and Muthukumar~\cite{Muthukumar}
suggested a linear dependence $\tau \sim N$ under a strong field. This is in 
agreement with some experimental results~\cite{Kasianowicz,Meller} for polymer 
translocation through $\alpha $-hemolysin channel. In addition, dynamic Monte 
Carlo (MC) simulation using Gaussian chain model~\cite{Chern} and Langevin dynamics 
simulation using rather short chains~\cite{Tian,Guo} seem to support the linear 
behavior. However, the above theories cannot explain the recent experimental 
result, namely that $\tau \sim N^{1.27}$ for polymer translocation through the 
solid-state nanopore ~\cite{Storm}. Recently, Kantor and Kardar~\cite{Kantor} 
demonstrated that the assumption of equilibrium in Brownian polymer dynamics 
by Sung and Park ~\cite{Sung} and Muthukumar~\cite{Muthukumar} breaks down more 
easily in the presence of a driving field and provided a lower bound $N^{1 + \nu }$ 
for the translocation time by 
comparison to the unimpeded motion of the polymer. However, in their simulations 
they failed to observe this scaling behavior, which was attributed to the finite 
size effect of the polymer length.~\cite{Kantor} In addition, they also checked a 
polymer that is being pulled by one end. A quadratic dependence of $\tau $ on
$N$ is predicted in their theory and confirmed by their 
simulations ~\cite{Kantor}.

Most recently, we have investigated both free~\cite{Luo1} and 
forced~\cite{Luo2} translocation using the two-dimensional fluctuating bond (FB) 
model with single-segment Monte Carlo moves. For the free translocation, 
to overcome the entropic barrier without artificial restrictions we considered 
a polymer, which is initially placed in the middle of the pore, and examined 
the escape time $\tau _e $ required for the polymer to completely exit the 
pore on either end. We found numerically that $\tau _e $ scales with the 
chain length $N$ as $\tau _e \sim N^{1 + 2\nu }$. This is the same scaling as 
predicted for the translocation time of a polymer which passes through the 
nanopore in one direction only ~\cite{Chuang,Milchev}. In addition, we also 
investigated the interplay between the pore length $L$ and the radius of gyration 
$R_{g}$. For $L \ll R_g $, we numerically verified that asymptotically 
$\tau _e \sim N^{1 + 2\nu }$, while for $L \gg R_g $, we found $\tau _e \sim N$. 
For forced translocation, we investigated the translocation dynamics under an external 
field within the pore. As our main result, we found a crossover scaling for 
the translocation time $\tau $ with the chain length from $\tau \sim N^{2\nu }$ for 
relatively short polymers to $\tau \sim N^{1 + \nu }$ for longer chains. Our 
results disagree with the experimental data that $\tau $ depends linearly on 
$N$ in the case of $\alpha $-hemolysin ~\cite{Kasianowicz,Meller},
but the predicted short chain 
exponent $2\nu \approx 1.18$ in three dimensions agrees reasonably well 
with the solid-state nanopore experiments of Storm \textit{et al.}~\cite{Storm}, 
where the experimental setup is closer to the theoretical models investigated.
No crossover in scaling behavior was observed in this experiment, presumably due 
to the fact that the polymers used in experiments are not long enough.

In this paper we investigate the translocation dynamics using Langevin 
dynamics simulations (LD). 
Previous LD simulations using rather short chains~\cite{Tian,Guo} 
seem to support the linear behavior for forced translocation time as a function of $N$, 
which is not expected to be correct for self-avoiding polymer. Also this result is in contradiction 
with the FB model simulations.~\cite{Luo2}
On the other hand, the FB model is based on single-segment Monte Carlo moves and thus neglects all 
translational degrees of freedom of the chain. Thus it is important to check that this does not 
adversely affect translocation dynamics.
In section \ref{chap-model}, we briefly describe our model and the simulation technique. 
In section \ref{chap-results}, we present our results for both free and driven translocation. 
For the free case, the emphasis is on verifying the scaling exponent obtained 
previously with the FB-model \cite{Kantor,Luo1}. For translocation under an electric field, 
we concentrate on studying the crossover in the scaling of translocation time as a function 
of the polymer length and on the influence of friction in the scaling exponent. 
Finally, the conclusions and discussion are in section \ref{chap-conclusions}.


\section{Model and method} \label{chap-model}

In the simulations, the polymer chains are modeled as bead-spring chains of Lennard-Jones (LJ) 
particles with the Finite Extension Nonlinear Elastic (FENE) potential. Both excluded volume and 
van der Waals interactions between beads are modeled by a repulsive LJ potential between all bead pairs: 
\begin{equation}
U_{LJ} (r)=\left\{  \begin{array}{ll}  4\epsilon \left[ \left(\frac{\sigma}{r}\right)^{12}-\left(\frac{\sigma}
{r}\right)^6  \right]+\epsilon, & r\le 2^{1/6}\sigma;\\
0, &  r>2^{1/6}\sigma,
\end{array}\right.
\end{equation}
where $\sigma$ is the diameter of a bead, and $\epsilon$ is the parameter adjusting the the depth 
of the potential. The connectivity between the beads is modeled as a FENE spring
\begin{equation}
U_{FENE} (r)=-\frac{1}{2}kR_0^2\ln(1-r^2/R_0^2),
\end{equation}
where $r$ is the separation between consecutive beads, $k$ is the spring constant and $R_0$ 
is the maximum allowed separation between connected beads. 

In the Langevin dynamics method, each bead is subjected 
to conservative, frictional, and random forces 
${\bf F}_i^C$, ${\bf F}_i^F$, and ${\bf F}_i^R$, respectively, 
and obeys the following equation of motion~\cite{Allen}
\begin{equation}
m{\bf \ddot {r}}_i = {\bf F}_i^C + {\bf F}_i^F + {\bf F}_i^R,
\label{Langevinequation}
\end{equation}
where $m$ is the monomer's mass. Excluded volume interaction is explicitly included in ${\bf F}_i^C$.
Hydrodynamic drag is included through the frictional force, which for individual monomers is 
${\bf F}_i^F=-\xi {\bf v}_i $, where $\xi$ is the friction coefficient, and ${\bf v}_i$ is the monomer's velocity.
The Brownian motion of the monomer resulting from the random bombardment of solvent molecules is included through
${\bf F}_i^R$ and can be calculated using the fluctuation dissipation theorem
\begin{equation}
  \begin{array}{ll} 
\langle {\bf F}_i^R (t) \rangle=0;    \\
\langle {\bf F}_i^R (t) \cdot {\bf F}_i^R (t') \rangle=6k_B T \xi \delta_{ij} \delta(t-t').
  \end{array}
\end{equation}
The conservative force in the Langevin equation consists of several
terms ${\bf F}_i^C={\bf F}_{LJ}+{\bf F}_{FENE}+{\bf F}_{\textrm{driving}}$. The driving force
depends on potential difference. For the spontaneous case there is no driving force.

The wall is described as $l$ columns of stationary particles within
distance $\sigma$ from one another and they interact with the beads
by the repulsive Lennard-Jones potential. Wall particle positions are
not changed in the simulations. The pore is introduced in the wall by
simply removing $w$ beads from each column.  
Under an electric field, the potential is expressed as
\begin{equation}
\label{eq1}
U_e = \left\{ {{\begin{array}{*{20}c}
 { -E \frac{l}{2},} \hfill \\
 { -Ex,} \hfill \\
 {E \frac{l}{2},} \hfill \\
\end{array} }} \right.\mbox{ }{\begin{array}{*{20}c}
 {x > l \mathord{\left/ {\vphantom {l 2}} \right. \kern-\nulldelimiterspace} 
2}; \hfill \\
 {l \mathord{\left/ {\vphantom {l 2}} \right. \kern-\nulldelimiterspace} 2 
\ge x \ge - l \mathord{\left/ {\vphantom {l {2\mbox{ and }y^2 \le \left( {w 
\mathord{\left/ {\vphantom {w 2}} \right. \kern-\nulldelimiterspace} 2} 
\right)^2}}} \right. \kern-\nulldelimiterspace} {2\mbox{ and }y^2 \le \left( 
{w \mathord{\left/ {\vphantom {w 2}} \right. \kern-\nulldelimiterspace} 2} 
\right)^2}}; \hfill \\
 {x < - l \mathord{\left/ {\vphantom {l 2}} \right. 
\kern-\nulldelimiterspace} 2}, \hfill \\
\end{array} }
\end{equation}
where $E$ is the electric field strength, $l$ is the pore length and $w$ is the pore width.

In our simulations, the LJ parameters $\epsilon$ and $\sigma$ fix the system energy 
and length units, respectively. Time scale is given by $t_{\textrm{LJ}}=(m\sigma^2/\epsilon)^{1/2}$.
The parameters are $\sigma=1$, $R_0=2\sigma$, $k=7\epsilon$. 
The Langevin equation is integrated in time by a method described by Ermak and 
Buckholtz \cite{Ermak} in 2D. For the pore, we set $w=2$ and $l=1$ 
unless otherwise stated. 
To create the initial configuration, the first monomer of the chain is placed in
the entrance of the pore. The polymer is then let to relax 
to obtain an equilibrium configuration such
that the first monomer position is fixed at the entrance but the other
monomers are under thermal collisions described by the Langevin thermostat. 
In all of our simulations we did a number of runs with uncorrelated initial
states and random numbers describing the random collisions. The estimate
for the translocation time was obtained by neglecting any failed
translocation and then calculating the average duration of the
accepted translocations. We note here that in some articles
e.g. in Ref.~\cite{Loebl} the estimate for the translocation time is chosen to be
the most probable translocation time. We checked that this does not change the scaling 
behavior of the chains.

\section{Results and discussion} \label{chap-results}
\subsection{Free translocation} \label{chap-spontaneous}

The translocation of a polymer through a nanopore faces a large entropic barrier due 
to the loss of a great number of available configurations. In previous simulations
~\cite{Chuang, Kantor}, an artificial restriction is imposed to prevent the first 
monomer from escaping from the pore. However, this restriction is unphysical.
Very recently, we suggested a new method to investigate the translocation dynamics without 
any restriction~\cite{Luo1}. 
In this method, the middle monomer is initially placed in the middle of the pore. 
The polymer can escape the pore from either side in time defined as the escape 
time $\tau_e$. Here, we use this method to investigate the distribution of escape times 
and the scaling of $\tau_e$ with the chain length. 
As discussed below, the distribution of  escape times is similar to the distribution of 
translocation time in the study by Chuang et al.~\cite{Chuang}. 
More importantly, escape time scales in the same way with the chain length as translocation time.


\begin{figure}
\begin{center} 
\includegraphics*[width=\figurewidth]{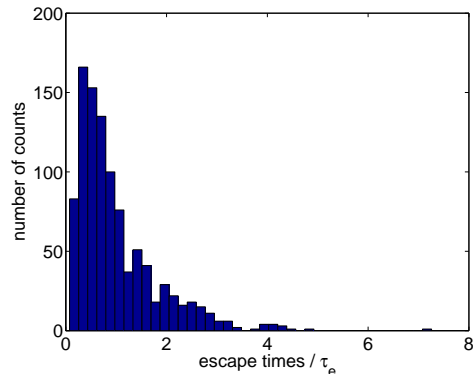}
\end{center}
\caption{
The distribution of 1000 escape times for a chain of length $N/2=80$ normalized 
by the mean value for free case. In the simulation, $\xi=0.7$ and $k_BT=1.2\epsilon$.
}
\label{distributionstwosided}
\end{figure}

\begin{figure}
\includegraphics*[width=\narrowfigurewidth]{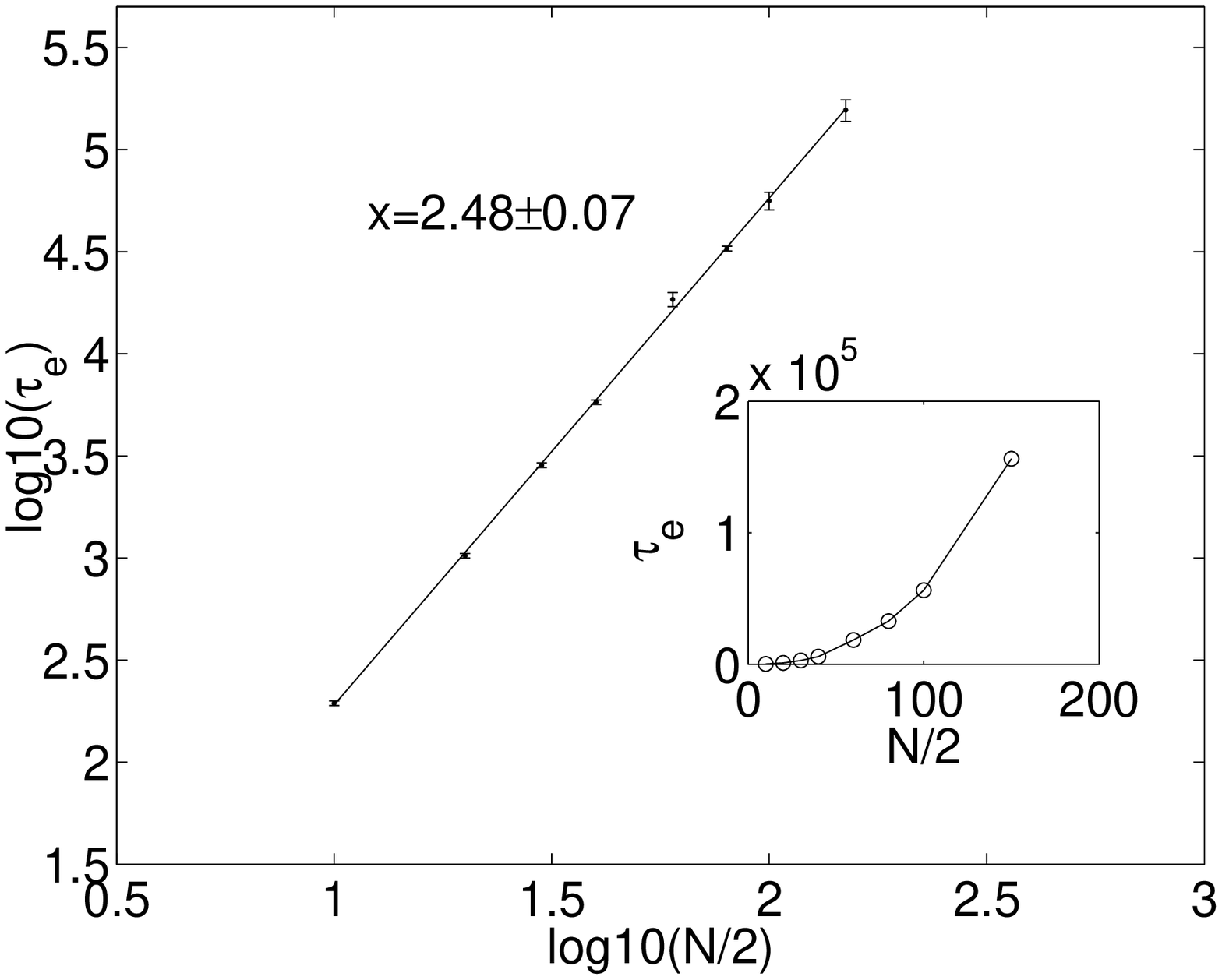}
\includegraphics*[width=\narrowfigurewidth]{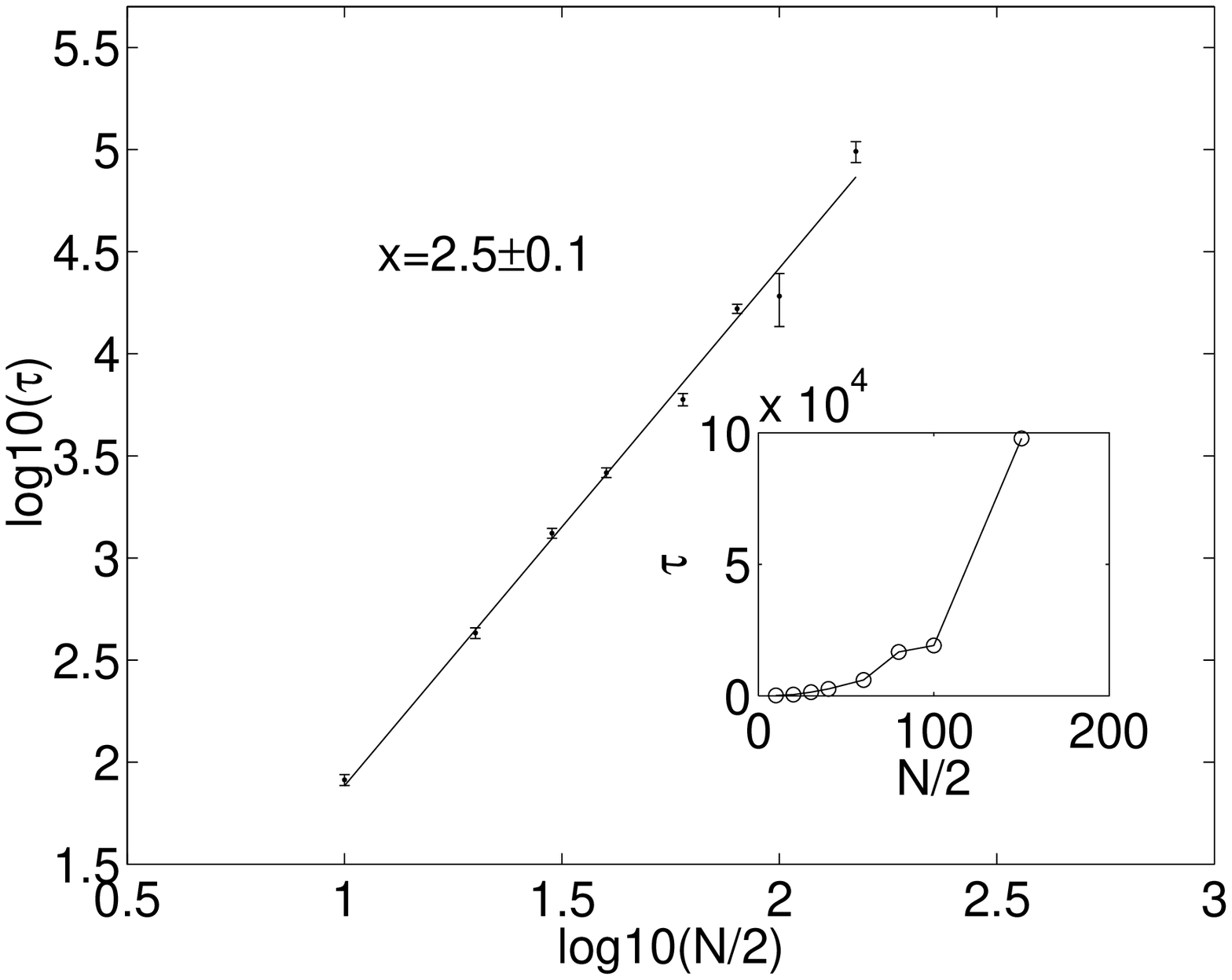}
{\hspace*{1.5cm}(a)} {\hspace*{3cm}(b)}
\caption{
(a) Average escape time $\tau_e$ and (b) the most probable escape time $\tau$
as a fuction of the polymer length for free case with $\xi=0.7$ 
and $k_BT=1.2\epsilon$. The averages where calculated over 1000 averages for chain lengths up to 
$N/2=80$. For chains of length $N/2=100 \textrm{ and }150 $ we had 300 and 100 averages, respectively.
}
\label{twosided}
\end{figure}

In the absence of the driving force translocation is extremely slow. 
For this reason we studied polymers of length only up to $N=300$. 
The distribution of escape times is shown in Fig. \ref{distributionstwosided} for a polymer 
of length $N/2=80$ normalized by its mean value.
We find that the escape times are distributed on a wide range with a long tail. 
The distribution is similar to those found previously with the FB model 
by Chuang {\it et al. }~\cite{Chuang}.
The average escape time $\tau_e$ as a function of $N$ is shown in Fig. \ref{twosided}a). 
We find the scaling exponent $x=2.48\pm 0.07 $. 
It is clear that average escape time is different from the most probable escape time. 
However, we have checked that the scaling exponents are almost the same. 
Using the most probable escape time, we find $x=2.5\pm 0.1 $, see Fig. \ref{twosided}b).
Above results are in excellent agreement with the 
theoretical prediction $\tau \sim N^{1+2\nu}$ ~\cite{Chuang,Milchev} and 
our previous results $\tau_e \sim N^{2.50\pm 0.01}$ ~\cite{Luo1}.

\subsection{Polymer translocation under an electric field}\label{chap-electric}

\begin{figure}
\includegraphics*[width=\narrowfigurewidth]{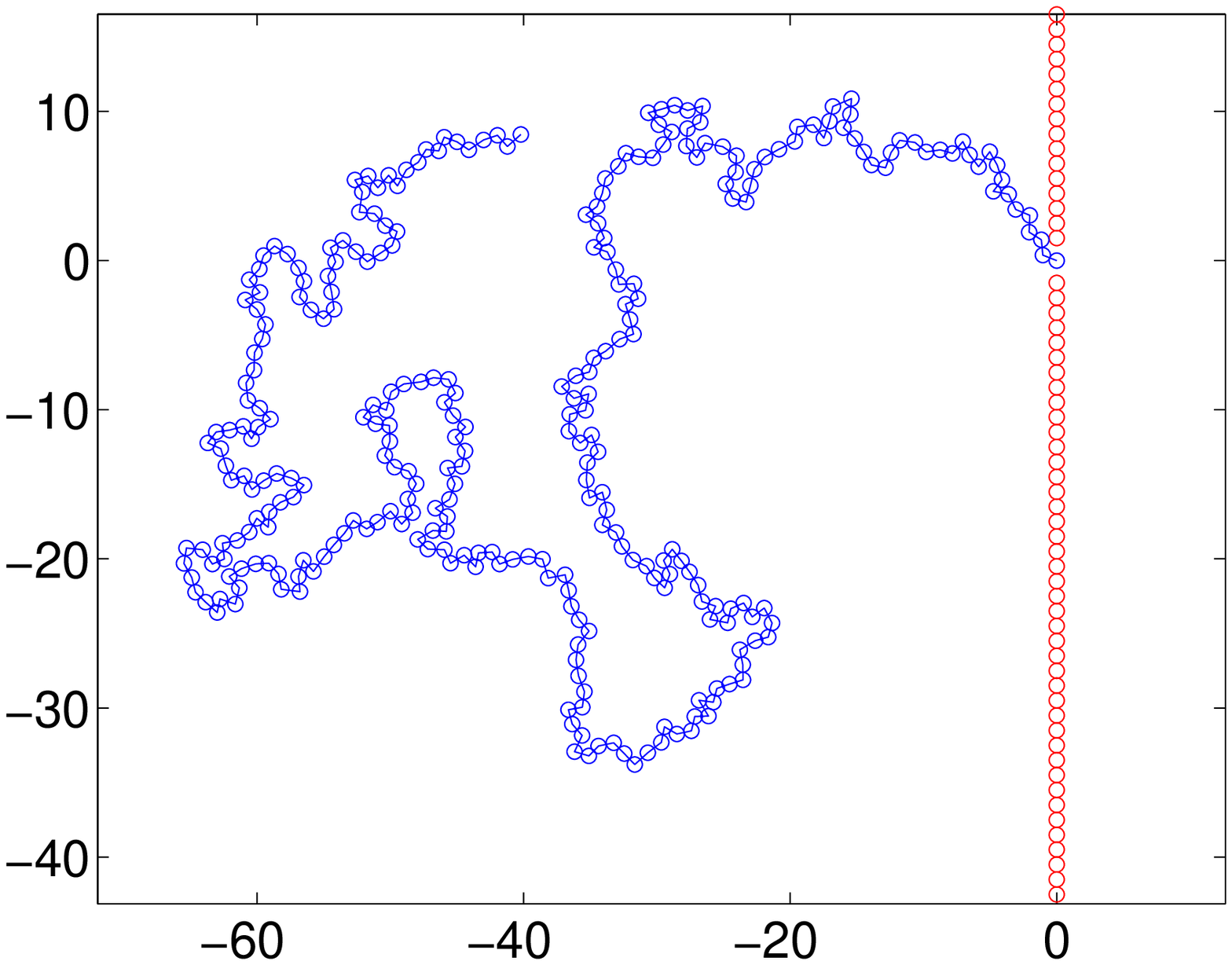}
\includegraphics*[width=\narrowfigurewidth]{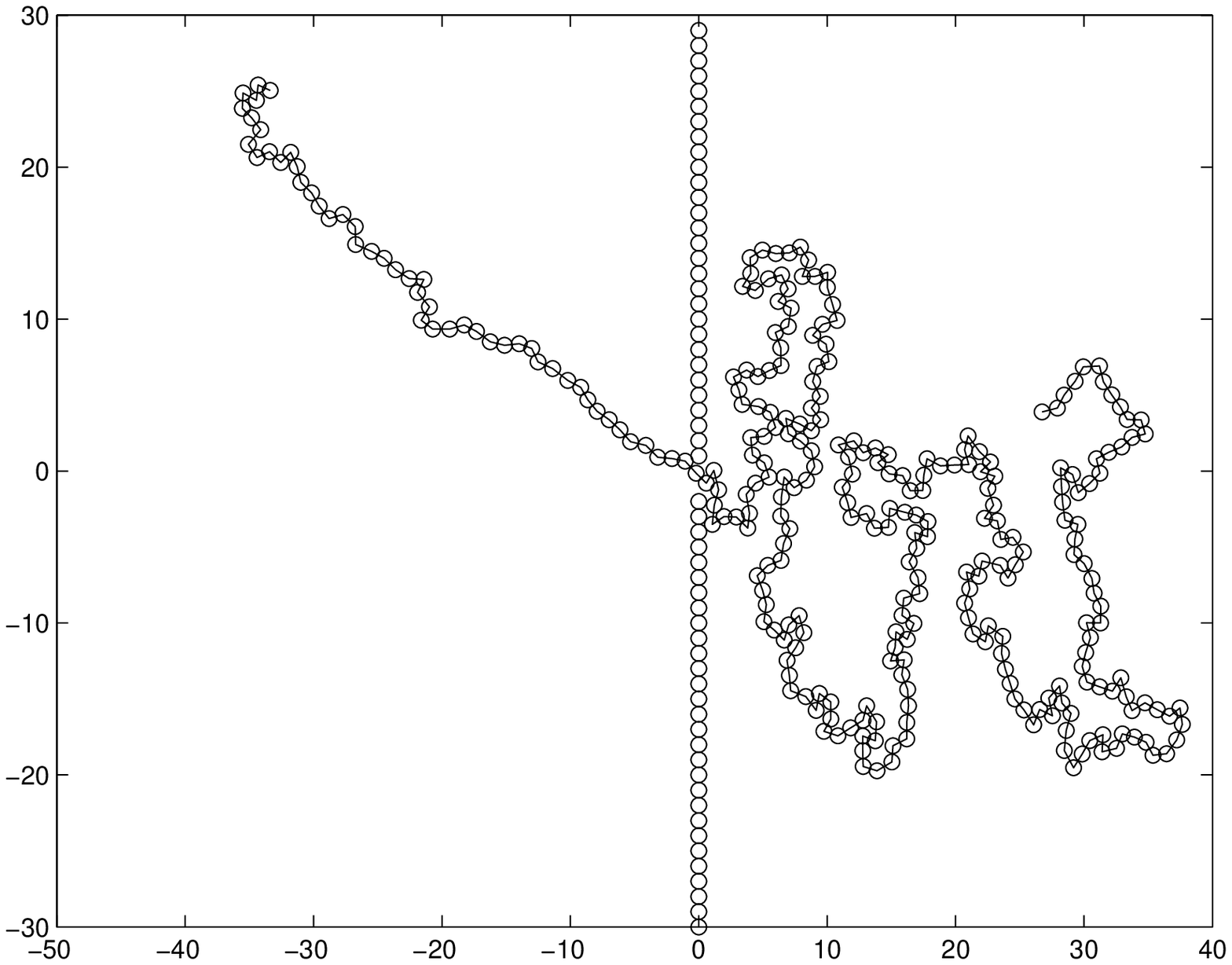}
{\hspace*{1.5cm}(a)} {\hspace*{3cm}(b)}
\caption{
The polymer configurations for $N=300$ under an electric field
$E=5\epsilon$, $\xi=0.7$ and $k_BT=1.2\epsilon$ 
before translocation (a) and during translocation (b). 
}
\label{confE}
\end{figure}

For driven translocation, the chain is initially placed on one side of the pore with one end of it 
in the pore entrance, as shown in Fig. \ref{confE}a). Then the chain is allowed to reach an 
equilibrium state, but with the constraint that the first monomer is fixed. Once the polymer is 
in its equilibrium state, the first monomer at the entrance of the pore is released, and that 
moment is designated as the start of the translocation. The translocation time is defined as 
the time duration taken for the chain to move through the pore. 
The typical snapshots of polymer configurations during translocation under an electric field is
presented in Fig. \ref{confE}b). 
It is important to note that not all simulation runs lead to a successful translocation and even when they do, 
translocation times  vary over a wide range of values. The success ratio 
describing the percentage of succesfull translocations depends on various factors. The driving 
force strength is the most important factor to increase the success probability. On the contrary, 
increasing the chain length and the frictional force, decreases the success probability. 
Below, we consider only the translocation time $\tau$ over all successful runs.

\subsubsection{Distribution of translocation times}

\begin{figure}
\includegraphics*[width=\figurewidth]{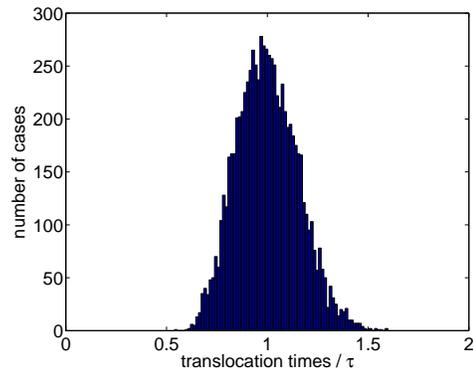}
\caption{
The distribution of 1000 translocation times for a chain of length $N=100$ normalized 
by the mean value in the case of an electric field of strength $E=5\epsilon$ as a driving force.
The other parameters are $\xi=0.7$ and $k_BT=1.2\epsilon$.
}
\label{distributionsE}
\end{figure}

As shown in Fig. \ref{distributionsE}, the distribution of translocation times has a qualitatively 
different shape compared with the distribution of the escape times in Fig. \ref{distributionstwosided} 
and the the distribution of translocation times studied by chuang et al.~\cite{Chuang} for the free case. 
In this case the distribution of translocation times is symmetric and narrow without a long tail. 
The stronger the diving force, the narrower distribution of translocation times.
A very important feature is that the average translocation time and the most probable translocation time 
are almost identical. Thus the average translocation time $\tau$ is well defined. 

\subsubsection{Effects of $\xi$ and $E$ on $\tau$}
In the Langevin equation, the friction coefficient $\xi$ describes the strength of coupling to the solvent. 
The value $\xi=0$ corresponds to the absence of the solvent molecules and would result in ballistic 
movement of particles. In the opposite limit $\xi \to \infty $, i.e. the overdamped limit, 
inertia plays no role in the dynamics.

\begin{figure}
\includegraphics*[width=\figurewidth]{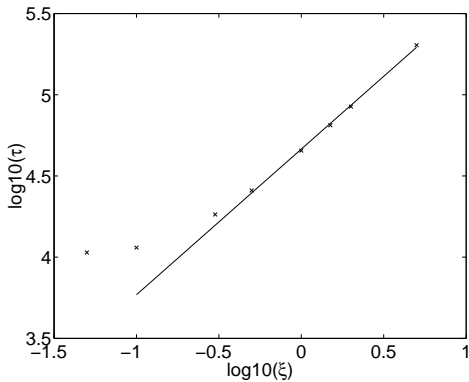}
\caption{
The translocation time as a function of the friction for a chain of length $N=100$ under an electric 
field $E=5\epsilon$ and $k_BT=1.2\epsilon$. $\tau$ is an average of 1000 translocations. The slope of the straight 
line in the figure is 1.
}
\label{tfric}
\end{figure}

The translocation time as a function of friction $\xi$ for $E=5\epsilon$ and $k_BT=1.2\epsilon$ 
is shown in Fig. \ref{tfric}.
As expected, $\tau$ increases with increasing $\xi$ due to the increases of the frictional force.
An important feature is that two regimes are observed. In the first rigime, when
$\xi < 0.5$, the translocation time increases very slowly with $\xi$. However, in the second regime, for  
$\xi > 0.5$, it increases rapidly and linearly with $\xi$. 
These results can be understood from the interplay between the driving force and the frictional force
according to the Langevin equation. For small $\xi$, the driving force and inertia play dominant roles 
compared with the frictional force.
As a result, the translocation time $\tau$ has a weak dependence on friction, thus increases only slightly with $\xi$. 
However, when $\xi$ is large enough, inertia can be neglected. According to the balance of the driving force and the 
frictional force, the horizontal velocity of the center of mass $v_{cmx} \sim \frac{E} {\xi}$. 
Thus the translocation time $\tau \sim \frac{1}{v_{cmx}} \sim \frac{\xi}{E}$ ~\cite{Footnote}. 
 
\begin{figure}
\includegraphics*[width=\figurewidth]{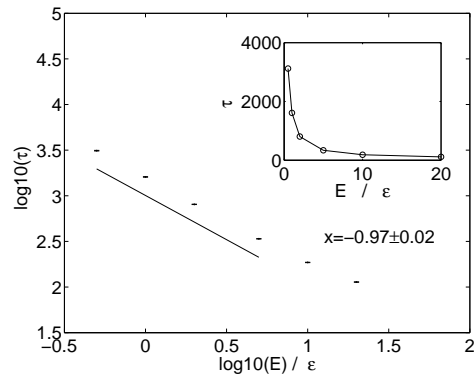}
\caption{
The translocation time as a function of electric field strength for a chain of length $N=100$. 
$\tau$ is an average of 1000 averages. Here $k_BT=1.2\epsilon$ and $\xi=0.7$.
} 
\label{tE}
\end{figure}

The influence of the driving force on the translocation time has also been investigated and the results are 
illustrated in Fig. \ref{tE} for a chain of length $N=100$.
We find that $\tau \sim E^{-0.97 \pm 0.02 }$ which is in excellent agreement with simple argument 
above that $\tau \sim E^{-1}$ in the regime where inertia can be neglected.
This finding is in excellent agreement with the results from the experiments of 
Kasianowicz {\it et al.} \cite{Kasianowicz}, who found that 
the translocation is inversely proportional to the electric field strength.
By contrast, our result disagrees with another experimental finding of 
an inverse quadratic dependence of $E$ for $\alpha $-hemolysin channel ~\cite{Meller}.

\subsubsection{Crossover scaling behavior of the translocation time with the chain length}

\begin{figure}
\includegraphics*[width=\narrowfigurewidth]{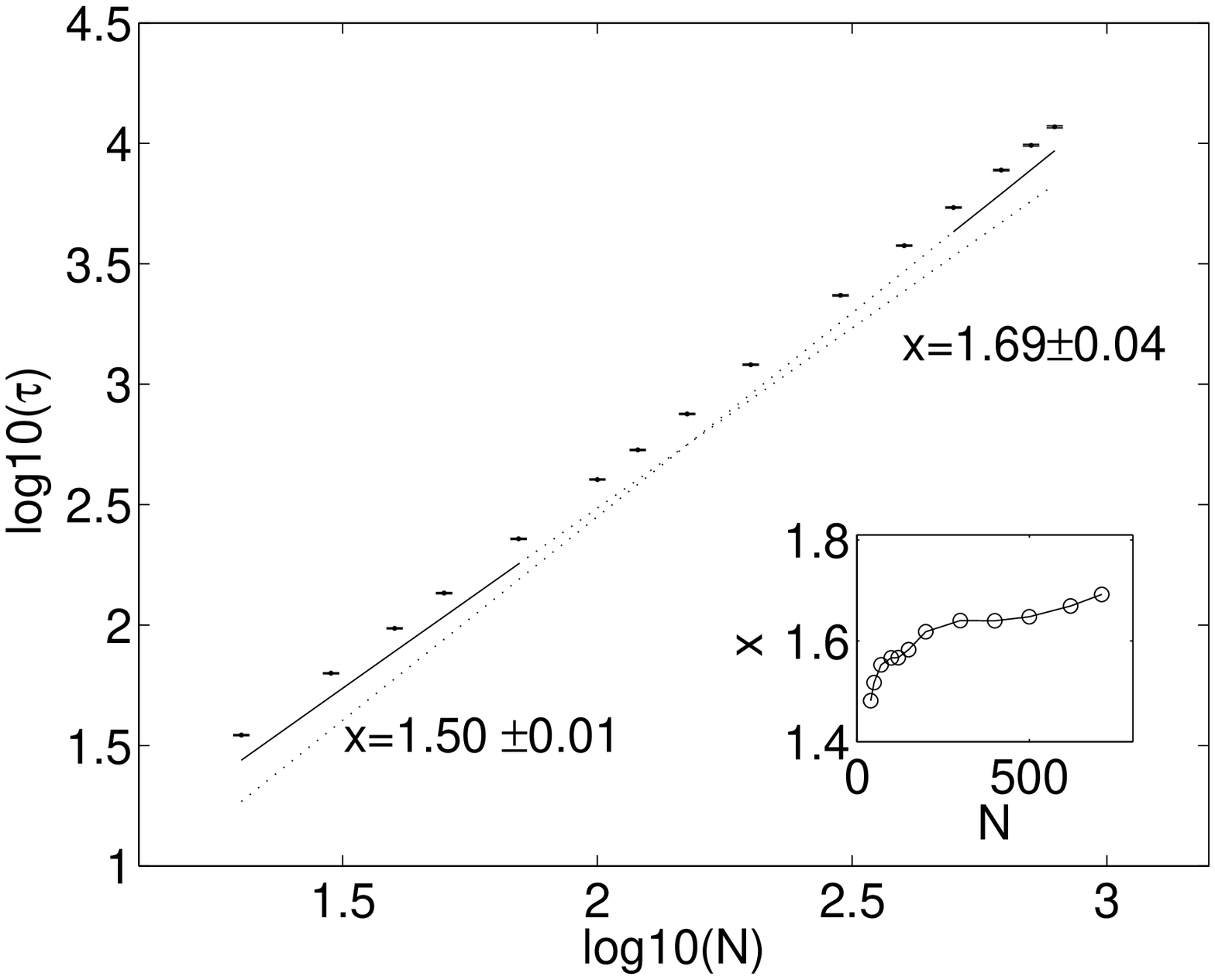}
\includegraphics*[width=\narrowfigurewidth]{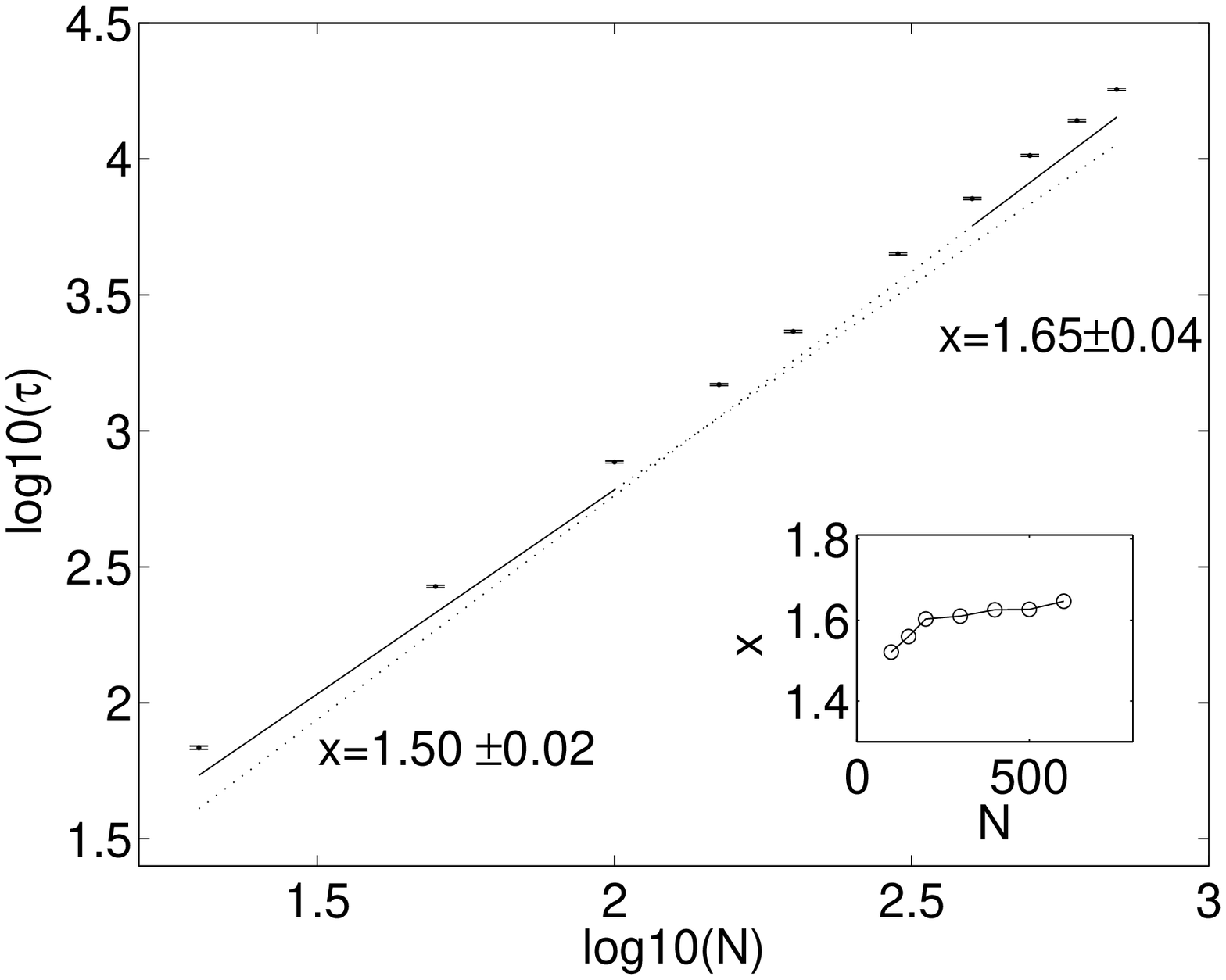}
{\hspace*{1.5cm}(a)} {\hspace*{3cm}(b)}
\includegraphics*[width=\narrowfigurewidth]{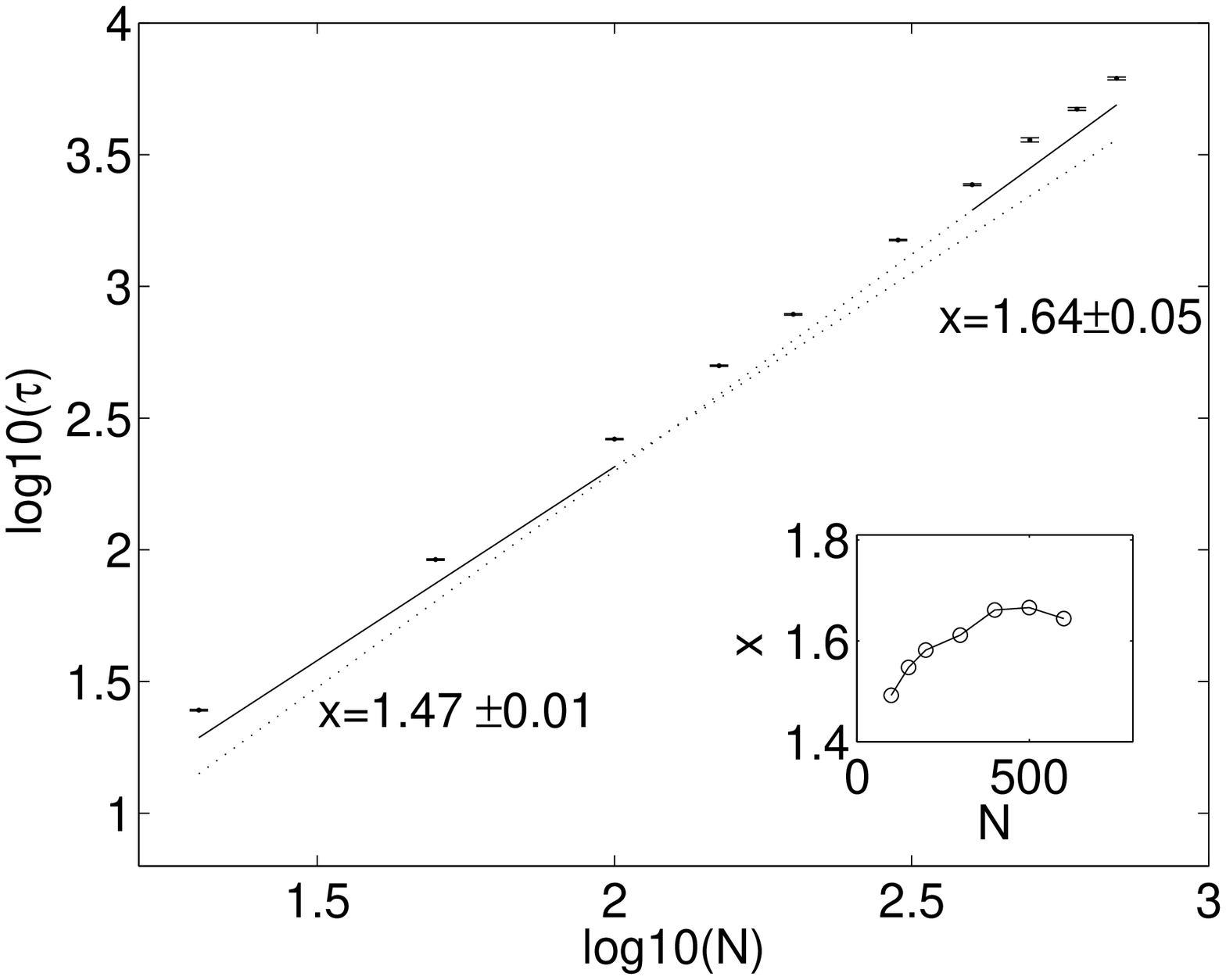}
\includegraphics*[width=\narrowfigurewidth]{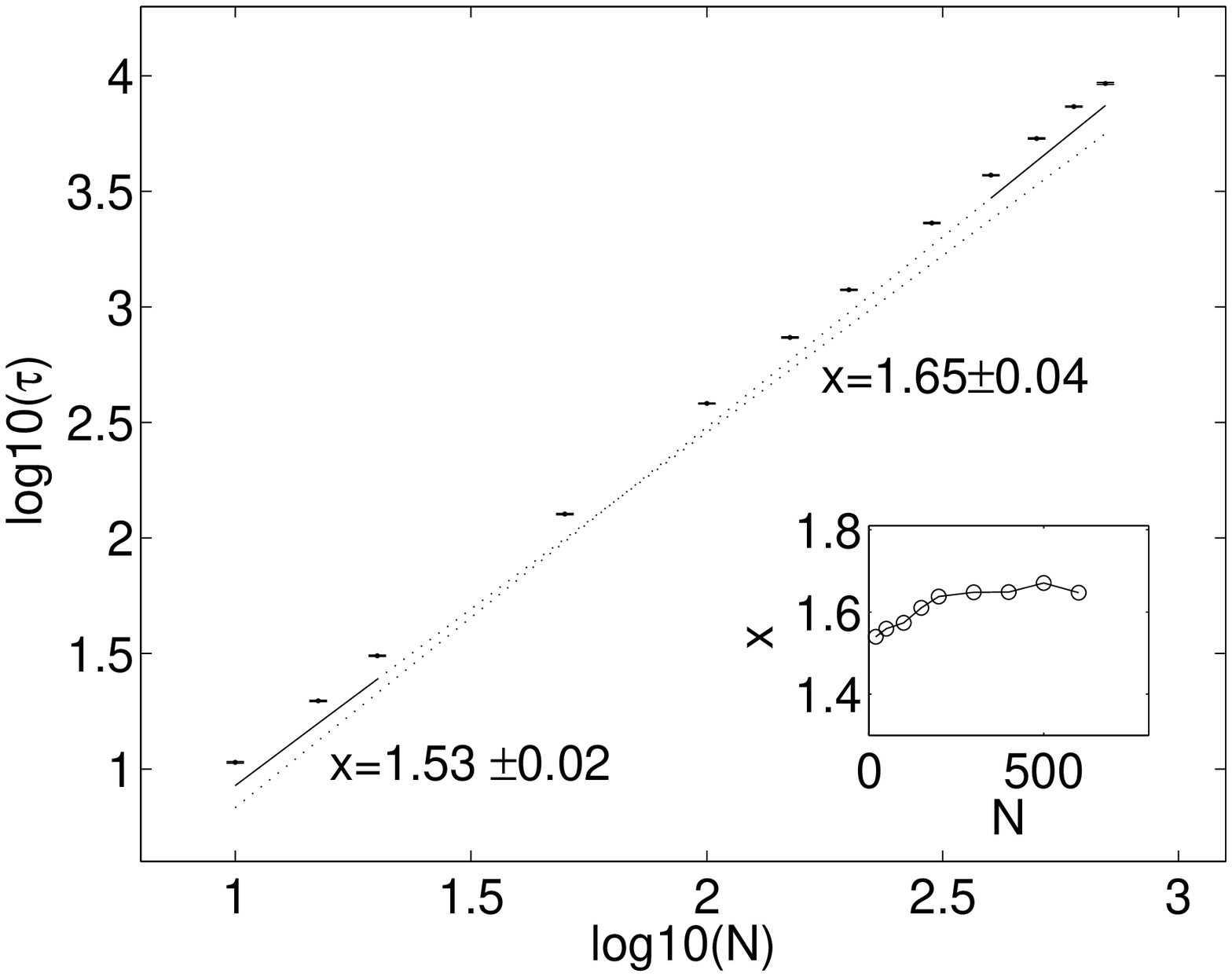}
{\hspace*{1.5cm}(c)} {\hspace*{3cm}(d)}
\caption{
The translocation time as a function of polymer length $N$ for $\xi=0.7$,
(a) $E=5\epsilon$, and $k_BT=1.2\epsilon$, 
(b) $E=2.4\epsilon$, and $k_BT=1.2\epsilon$,
(c) $E=8.3\epsilon$, and $k_BT=2\epsilon$,
(d) $E=5\epsilon$, $k_BT=1.2\epsilon$, and $w=3$.
$\tau$ is an average of 1000 runs. In each figure, a shifted solid line is plotted beneath the data to which 
the curve is fitted. The curve is continued with the same slope but for clarity it is plotted as a dashed line. 
The inset shows the local scaling exponent $x$ in which each point is the slope of a curve fitted to four points.
}
\label{tNE}
\end{figure}

The translocation time as a function of the polymer length with $\xi=0.7$ is presented in Fig. \ref{tNE}. 
In each figure, a shifted solid line is plotted beneath the data to which the curve is fitted. 
The curve is continued with the same slope but for clarity it is plotted as a dashed line. 
The inset shows the local scaling exponent $x$ in which each point is the slope of a curve fitted to four points.

\begin{figure}
\includegraphics*[width=\narrowfigurewidth]{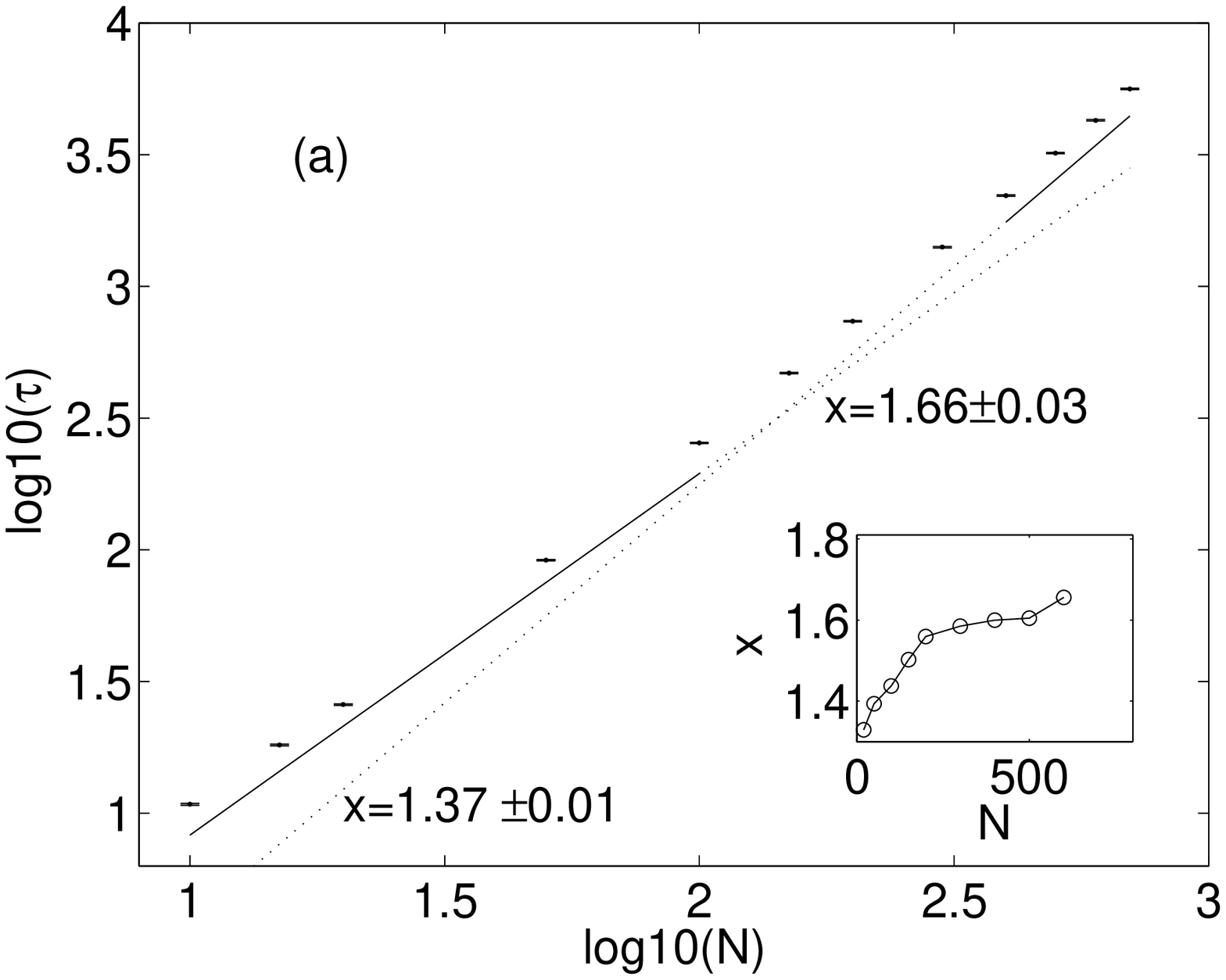}
\includegraphics*[width=\narrowfigurewidth]{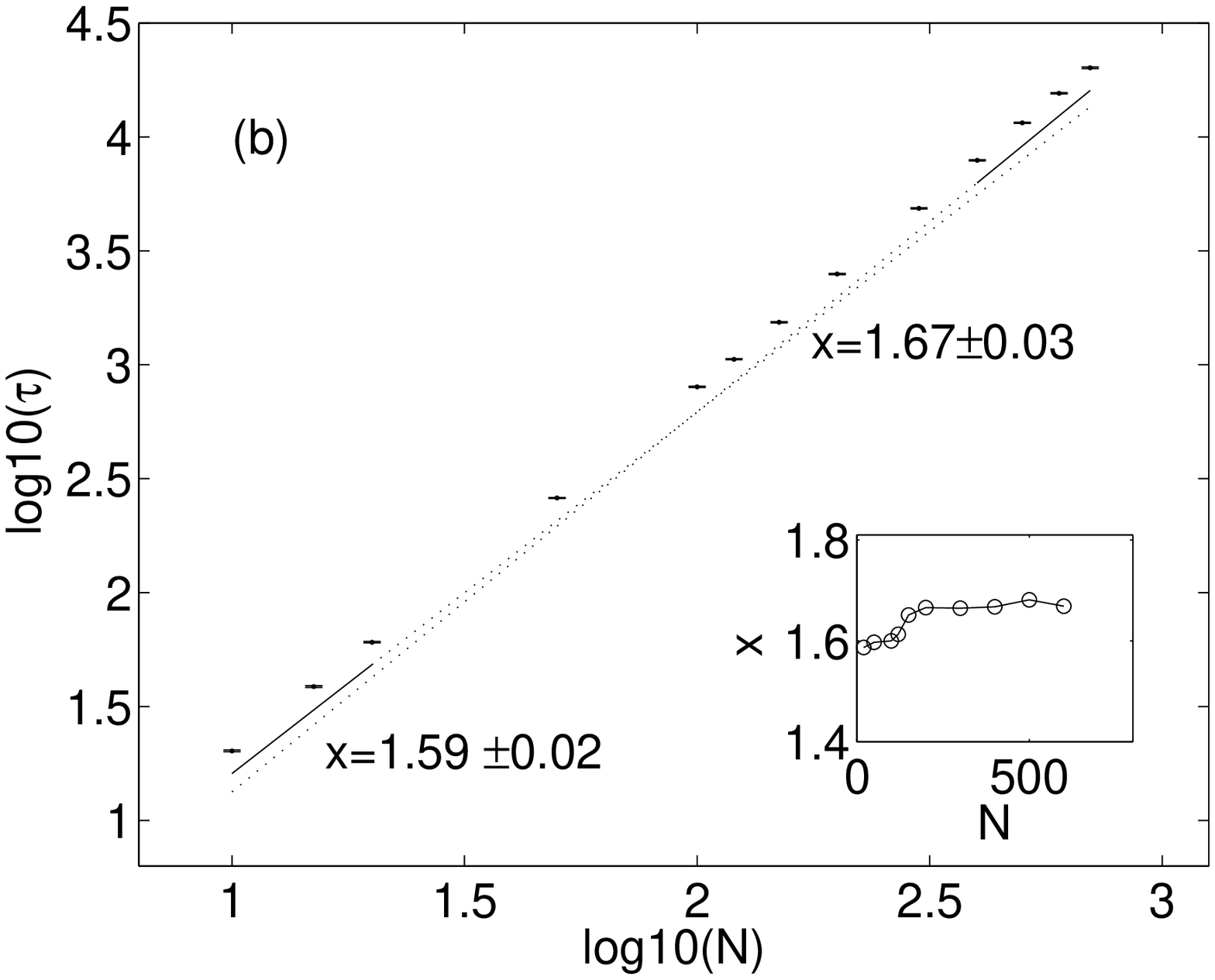}
{\hspace*{1.5cm}(a)} {\hspace*{3cm}(b)}
\includegraphics*[width=\narrowfigurewidth]{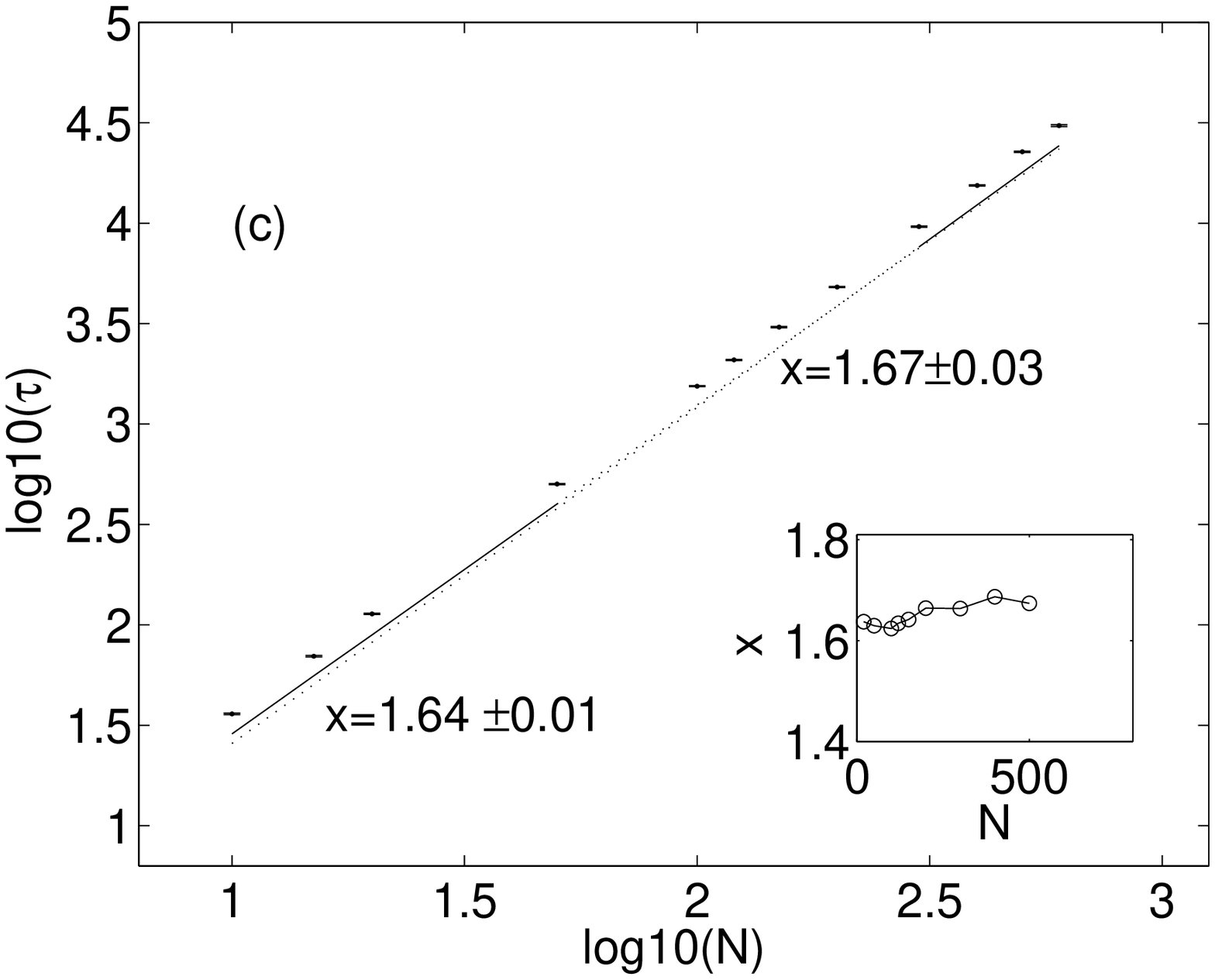}
{\hspace*{1.5cm}(c)} 
\caption{
The translocation time as a function of polymer length $N$ with an electric field of 
strength $E=5\epsilon$ as a driving force for $k_BT=1.2\epsilon$ and (a) $\xi=0.4$, (b) $\xi=1.5$ and (c) $\xi=3$.
$\tau$ is an average of 1000 runs.
}
\label{tNfric0415}
\end{figure}

The result for $E=5\epsilon$ and $k_BT=1.2\epsilon$ is shown in Fig. \ref{tNE}a). One of the main features 
is that a crossover scaling behavior is observed. 
For short chains ($N \le 70$), the scaling exponent $x=1.50\pm 0.01$, which is in good agreement 
with 2$\nu $, where $\nu =0.75$ is the Flory exponent for a self-avoiding walk in 2D. 
However, for $N\ge 300$ the exponent becomes $x=1.69\pm 0.04$, which is close to $1+\nu = 1.75$.
The polymer assumes a more compact configuration resulting from both the dynamical effect and geometrical 
restriction. This leads to a smaller effective value for the exponent $\nu$ as compared to its Flory 
value ~\cite{Bhatt}.
For the different electric strength $E$ and temperature $T$ and pore width $w$,
the same exponent and crossover behavior is observed, as shown in Fig. \ref{tNE}b), c) and d), respectively.
Next, we investigate the effect of the friction on the scaling exponent. In Fig. \ref{tNfric0415} we 
present results for $\tau$ as a function of $N$ with $E=5\epsilon$ for three different frictions: 
$\xi=0.4$, $\xi=1.5$ and $\xi=3$. 
For $\xi=0.4$ and $\xi=1.5$, a clear crossover is observed, but for $\xi=3$ only 
$\tau \sim N^{1+2\nu}$ occurs.

To understand the above results, one should consider the configuration of the monomers after exiting 
from the during pore under 
an electric field, as shown in Fig. \ref{confE}b). 
For driven translocation, there is a qualitative difference compared with the purely diffusive free 
translocation for long chains: translocation under a strong driving force is much faster 
than that for free diffusion and thus the translocated monomers don't have time to diffuse 
away from the vicinity of the pore exit, see Fig. \ref{confE}b).
As to the effect of the friction on the translocation dynamics, 
on the one hand, increasing $\xi$ leads to the increase of the frictional force, resulting in 
slowing down of the translocation. 
On the other hand, the Rouse relaxation time for the self-avoiding chain is proportional to 
$\frac {\xi}{k_BT}$, which rapidly increases with $\xi$. The latter factor is always dominant 
and thus the translocation time is much shorter than the Rouse relaxation time. 
As a result,it is more difficult for translocated monomers to diffuse away from
the pore exit even for short chains with higher $\xi$. We checked that the radius of gyration 
is larger before translocation than that immediately after it. This fact indicates the higher density of 
translocated monomers near the pore exit.
It can be imagined that the higher density of translocated monomers near the pore exit greatly slows down 
the translocation.
Therefore, for intermediate values of $\xi$, two scaling regimes for short and long chains are observed, respectively.
However, for high friction such as $\xi=3$, the Rouse relaxation time 
is so long compared to the translocation time that it is more difficult for the translocated monomers 
to diffuse away even for short chains, thus the translocation dynamics directly 
enters into the regime where $\tau \sim N^{1+2\nu}$.

Our results disagree with the experimental data that $\tau $ depends linearly 
on $N$ for translocation through $\alpha $-hemolysin pore ~\cite{Kasianowicz,Meller}, but the predicted short 
chain exponent $2\nu = 1.18$ in 3D agrees well with the 
solid-state nanopore experiments of Storm \textit{et al.}~\cite{Storm}, who found an exponent of 
1.27. As pointed out by Storm \textit{et al.}~\cite{Storm}, the linear behavior for $\alpha $-hemolysin pore 
may be from the specific and complicated interaction between DNA chain and the pore.
The beginning of the crossover region occurs at $N\ge 300$ which 
is beyond or near the longest ssDNA and dsDNA used in the experiments so far ~\cite{Kasianowicz,Meller,Storm}. 
Thus, it is not surprising that crossover in scaling behavior has not been experimentally observed yet.
Theoretically, as expected our results disagree with the linear dependence $\tau \sim N$ prediction 
by Sung and Park~\cite{Sung} and Muthukumar~\cite{Muthukumar}, which is invalid for nonequilibrium translocation. 
However, our results also disagree with the previous Langevin dynamics simulations 
for relatively short polymers ~\cite{Tian, Loebl}, which show $\tau \sim N$. 
The present results agree with the main findings in our previous FB-studies \cite{Luo2} for 
both short and long chains and the theoretical prediction $\tau \sim N^{1+\nu}$ 
by Kantor and Kardar \cite{Kantor}. 

\subsubsection{Crossover behavior of the translocation velocity}

To study translocation dynamics in detail, we also calculated the translocation velocity $v$ as a 
function of polymer length $N$. The translocation velocity can be measured in several ways. 
A simple way is to measure the average horizontal velocity of the 
center of mass of the polymer over the whole duration of all successful runs
\begin{equation}
v = \left\langle {v_{cmx} } \right\rangle = \left\langle 
{\frac{1}{N}\sum\limits_{i = 1}^N {v_{xi} } } \right\rangle ,
\label{vcmx}
\end{equation}
where $v_{xi}$ is the horizontal component of the velocity of monomer $i$. 
In addition, the translocation velocity can also be defined as
\begin{equation}
v = \frac{\left\langle {x_0 } \right\rangle }{\left\langle {\tau 
_i } \right\rangle }.
\label{v0}
\end{equation}
where $x_0$ is the horizontal coordinate of the last monomer in the initial equilibrated configuration 
of the polymer before the translocation and $\tau_i$ is the translocation time for every successful run.

According to our numerical results for crossover scaling behavior, we must have that
\begin{equation}
\label{vcrossover}
v \sim \frac{N^\nu }{\tau } \sim \left\{ {{\begin{array}{*{20}c}
 {\frac{N^\nu }{N^{2\nu }} \sim N^{ - \nu }\mbox{ for small }N} \hfill \\
 {\frac{N^\nu }{N^{1 + \nu }} \sim N^{ - 1}\mbox{ for large }N} \hfill \\
\end{array} }} \right..
\end{equation}

\begin{figure}
\includegraphics*[width=\narrowfigurewidth]{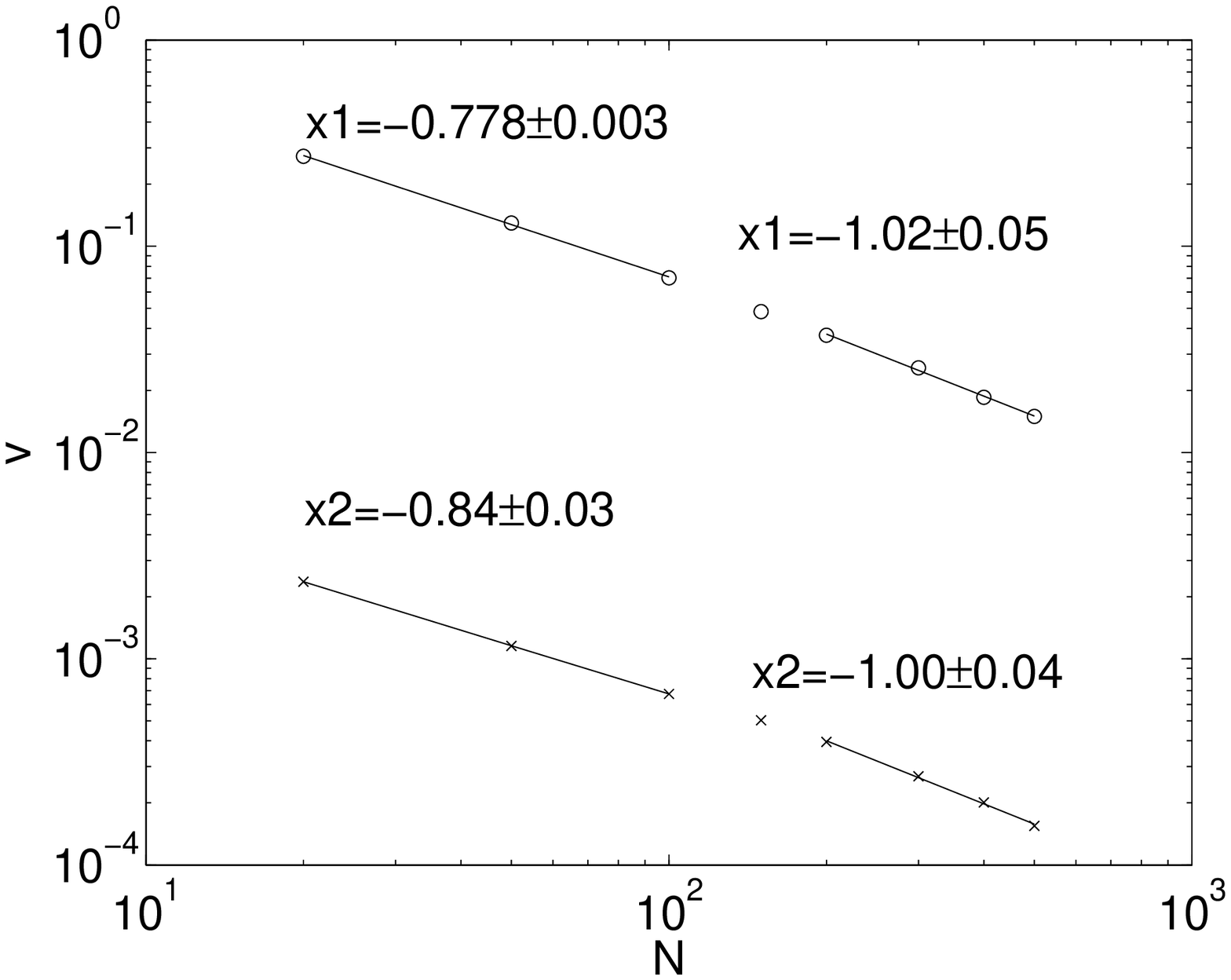}
\includegraphics*[width=\narrowfigurewidth]{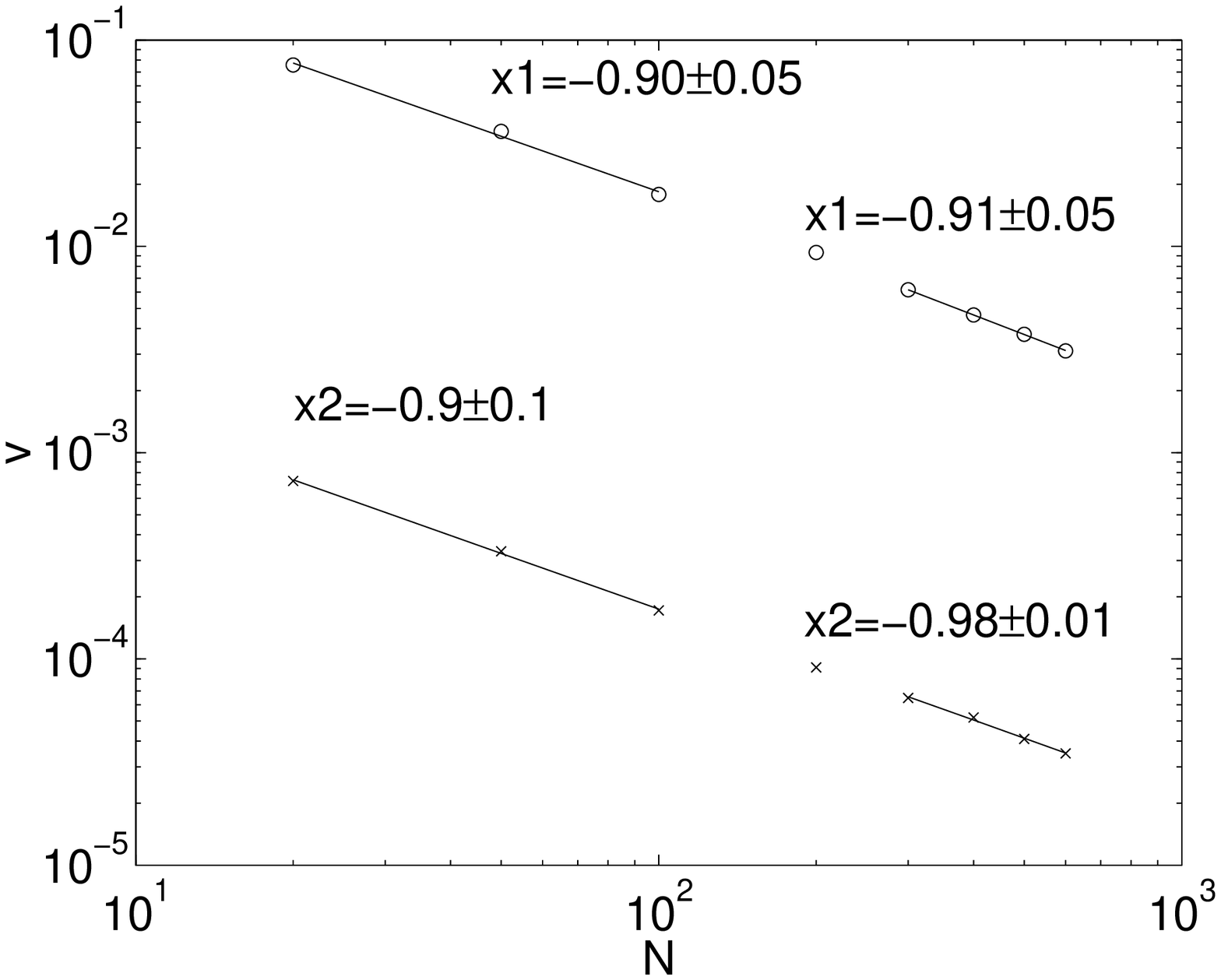}
{\hspace*{1.5cm}(a)} {\hspace*{3cm}(b)}
\caption{
The translocation velocity as a function of polymer length $N$ with an electric field of 
strength $E=5\epsilon$, $k_BT=1.2\epsilon$ and (a) $\xi=0.7$ and (b) $\xi=3$. 
Averages are taken over 100 runs. 
$x1$ corresponds to the velocities defined as in Eq.(\ref{v0}) and the $x2$ to velcities defined as 
in Eq.(\ref{vcmx}). The latter results have been shifted for clarity.
}
\label{vNE5}
\end{figure}

In Fig. \ref{vNE5}a), we present the polymer velocity as a function of chain length for 
$E=5\epsilon$, $k_BT=1.2\epsilon$ and $\xi=0.7$.
We get $v \sim N^{-0.778\pm 0.003}$ and $v \sim N^{-0.842\pm0.03} $ for short chains 
and $v \sim N^{-1.02\pm 0.05}$ and $v\sim N^{-1.00\pm0.04}$ for long chains, respectively. 
There is a clear crossover in the translocation velocity from $v \sim N^{-\nu}$ to $v\sim N^{-1}$, 
which is in agreement with Eq.(\ref{vcrossover}). This simple test confirms that the crossover in the 
translocation time takes place because of a crossover in the translocation velocity. 
Here, we should mention that for high friction such as $\xi=3$, there is no crossover 
and only $v\sim N^{-1}$ is observed, see Fig. \ref{vNE5}b). 


\subsection{Waiting times}\label{chap-waiting}

\begin{figure}
\includegraphics*[width=\narrowfigurewidth]{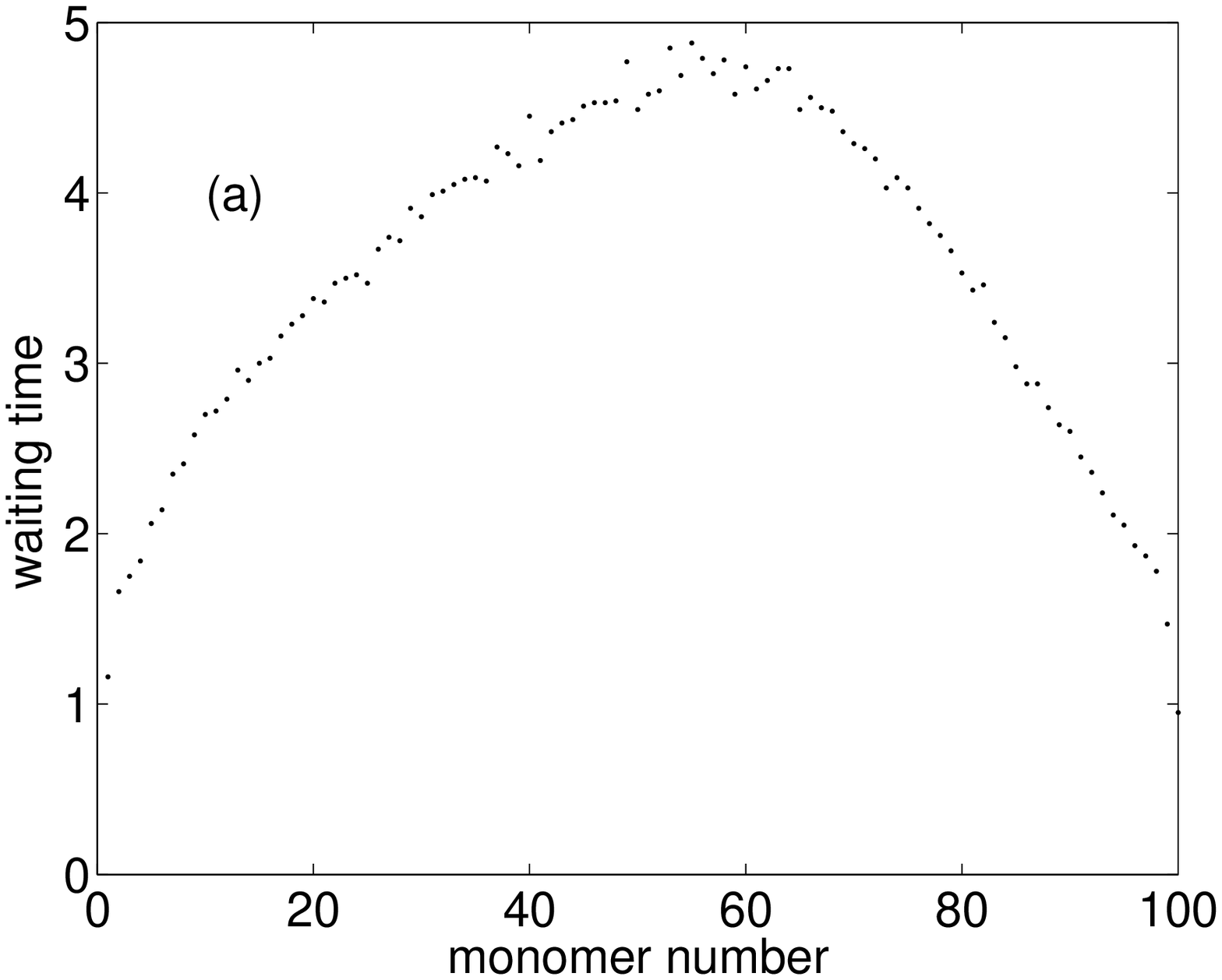}
\includegraphics*[width=\narrowfigurewidth]{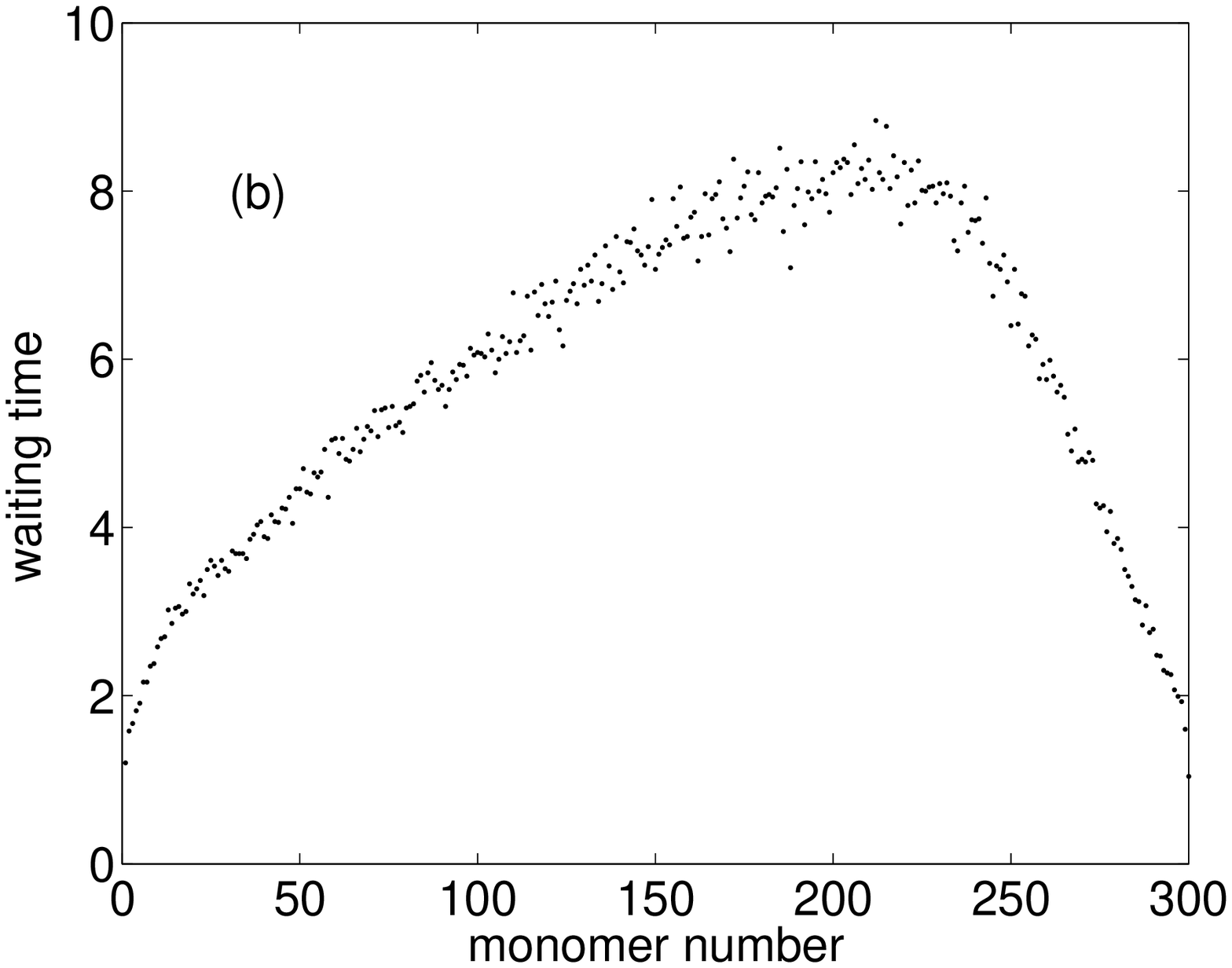}
{\hspace*{1.5cm}(a)} {\hspace*{3cm}(b)}
\caption{
The waiting times for a polymer of (a) length $N=100$, (b) $N=300$ for 
$E=5\epsilon$, $k_BT=1.2\epsilon$ and $\xi=0.7$. The waiting times are averages of 1000 runs.
}
\label{waitingN100}
\end{figure}

The dynamics of a single-segment passing through the pore during translocation is an important issue, 
as far as experiments are concerned. The non-equilibrium nature of translocation has a considerable effect 
on this. We have numerically calculated the waiting times for each monomer passing through the pore. 
We define the waiting time of monomer $s$ as the average time between the events that monomer $s$ and 
monomer $s+1$ exit the pore. 
The waiting times for chains of length $N=100$ and $N=300$ are presented in Fig. \ref{waitingN100}.
 
The waiting time depends strongly on the monomer position in the chain. For the shorter polymer 
$N=100$ the monomers approximately in the middle of the polymer need the longest time to translocate and the 
distribution is close to symmetric. However, for the polymer of length $N=300$, it's approximately 
the 220$^{\textrm{th}}$ monomer that needs the longest time to thread the pore, which 
indicates that during late stages of translocation the high density of 
segments of a long polymer near the pore slows down the translocation. 
The same phenomenon was found in our earlier study with the FB-model \cite{Luo2}.

\section{Conclusions} \label{chap-conclusions}
In this work, we investigate the dynamics of polymer translocation 
through a nanopore using the two-dimensional (2D) Langevin dynamics simulations.
For free translocation, we consider a polymer which is initially placed
in the middle of the pore and study the escape time $\tau_e$. 
The distribution of the escape times is wide 
and has a long tail. We verified that average escape time (as well as the most probable escape time) 
scales with the chain length $N$ as $\tau_e \sim  N^{1+2\nu}$. 
These results confirm previous theoretical studies based on a less miscrosopic fluctuating bond model.
For forced translocation, we concentrate 
on studying the influence of the friction $\xi$, driving force $E$ and polymer chain length $N$. 
For strong driving forces, the distribution of translocation 
times is narrow without a long tail and symmetric, which is completely different compared 
with free translocation.
The influence of the $\xi$ on the translocation time depends
on the ratio between driving force and frictional force. 
We find translocation time $\tau$ is inverse proportional to driving force for a wide range of friction values. 
Finally, as regards to the dependence of translocation on the length of the polymer,
 we find a crossover scaling for the $\tau$ with $N$ from $\tau \sim 
N^{2\nu }$ for relatively short polymers to $\tau \sim N^{1 + \nu }$ for 
longer chains at moderate values of the friction. 
For higher $\xi$, there is no crossover because in this case the Rouse
relaxation time is extremely long compared with translocation time.
These scaling behaviors can be understood from the observation that in the limit of large 
$N$ and $\xi$, there is a high density of segments 
near the exit of the pore, which slows down the translocation process due to 
slow relaxation of the chain.

\begin{acknowledgments}
This work has been supported in part by a 
Center of Excellence grant from the Academy of Finland (COMP). We also wish to 
thank the Center for Scientific Computing Ltd. for allocation of computer 
time. 
\end{acknowledgments}



\begin{thebibliography}{10}

\bibitem{Alberts} B. Alberts and D. Bray, \textit{Molecular Biology of the Cell }(Garland, New York, 1994).
\bibitem{Darnell} J. Darnell, H. Lodish, and D. Baltimore, \textit{Molecular Cell Biology }
                 (Scientific American Books, New York, 1995).
\bibitem{Miller} R. V. Miller, \textit{Sci. Amer.} {\bf 278}, 66 (1998).
\bibitem{Han} J. Han, S. W. Turner, and H. G. Craighead, \textit{Phys. Rev. Lett.} {\bf 83}, 1688 (1999).
\bibitem{Turner} S. W. P. Turner, M. Calodi, and H. G. Craighead, \textit{Phys. Rev. Lett.} {\bf 88}, 128103 (2002).
\bibitem{Chang}  D.-C. Chang, \textit{Guide to Electroporation and Electrofusion }(Academic, New York, 1992).
\bibitem{Kasianowicz} J. J. Kasianowiczs, E. Brandin, D. Branton and D. W. Deaner, 
							\textit{Proc. Natl. Acad. Sci. U.S.A.} {\bf 93}, 13770 (1996).
\bibitem{Aktson} M. Aktson, D. Branton, J. J. Kasianowicz, E. Brandin, and D. W. Deaner, 
                                          \textit{Biophys. J.} {\bf 77}, 3227 (1999).
\bibitem{Meller2} A. Meller, L. Nivon, E. Brandin, J. A. Golovchenko, and D. Branton, 
					\textit{Proc. Natl. Acad. Sci. U.S.A.} {\bf 97}, 1079 (2000).
\bibitem{Henrickson} S. E. Henrickson, M. Misakian, B. Robertson, and J. J. Kasianowicz, \textit{Phys. 
					Rev. Lett.} {\bf 85}, 3057 (2000).
\bibitem{Meller} A. Meller, L. Nivon, and D. Branton, \textit{Phys. Rev. Lett.} {\bf 86}, 3435 (2001).
\bibitem{Sauer} A. F. Sauer-Budge, J. A. Nyamwanda, D. K. Lubensky, and D. Branton, Phys. 
					Rev. Lett. {\bf 90}, 238101 (2003).
\bibitem{Meller3} A. Meller, \textit{J. Phys.: Condens. Matter} {\bf 15}, R581 (2003).
\bibitem{Li} J. L. Li, D. Stein, C. McMullan, D. Branton, M. J. Aziz, and J. A. 
					Golovchenko, \textit{Nature} (London) {\bf 412}, 166 (2001).
\bibitem{Gershow} J. L. Li, M. Gershow, D. Stein, E. Brandin, and J. A. Golovchenko, 
					\textit{Nat. Mater.} {\bf 2}, 611 (2003).
\bibitem{Chen} A. J. Storm, J. H. Chen, X. S. Ling, H. W. Zandbergen, and C. Dekker, 
					\textit{Nat. Mater.} {\bf 2}, 537 (2003).
\bibitem{Storm} A. J. Storm, C. Storm, J. Chen, H. Zandbergen, J. -F. Joanny and C. Dekker, 
							\textit{Nano Lett.} {\bf 5}, 1193 (2005).
\bibitem{Simon} S. M. Simon, C. S. Peskin, and G. F. Oster, \textit{Proc. Natl. Acad. Sci. U.S.A.} 
							{\bf 89}, 3770 (1992).
\bibitem{Sung}   W. Sung and P. J. Park, \textit{Phys. Rev. Lett.} {\bf 77}, 783 (1996).
\bibitem{Park} P. J. Park and W. Sung, \textit{J. Chem. Phys.}  {\bf 108}, 3013 (1998).
\bibitem{diMarzio} E. A. diMarzio and A. L. Mandell, \textit{J. Chem. Phys.}  {\bf 107}, 5510 (1997).
\bibitem{Muthukumar}   M. Muthukumar, \textit{J. Chem. Phys.} {\bf 111}, 10371 (1999). 
\bibitem{Muthu2} M. Muthukumar, \textit{J. Chem. Phys.}  {\bf 118}, 5174 (2003).
\bibitem{Lubensky} D. K. Lubensky and D. R. Nelson, \textit{Biophys. J.} {\bf 77}, 1824 (1999).
\bibitem{Slonkina} E. Slonkina and A. B. Kolomeisky, \textit{J. Chem. Phys.}  {\bf 118}, 7112 (2003).
\bibitem{Ambj} T. Ambjornsson, S. P. Apell, Z. Konkoli, E. A. DiMarzio, and J. J. 
				Kasianowicz, \textit{J. Chem. Phys.}  {\bf 117}, 4063 (2002).
\bibitem{Metzler} R. Metzler and J. Klafter, \textit{Biophys. J.} {\bf 85}, 2776 (2003).
\bibitem{Ambj2} T. Ambjornsson and R. Metzler, \textit{Phys. Biol.} {\bf 1}, 19 (2004).
\bibitem{Ambj3} T. Ambjornsson, M. A. Lomholt, and R. Metzler, \textit{J. Phys.: Condens. Matter} {\bf 17}, S3945 (2005).
\bibitem{Gerland} U. Gerland, R. Bundschuh, and T. Hwa, \textit{Phys. Biol.} {\bf 1}, 19 (2004).
\bibitem{Baumg} A. Baumgartner and J. Skolnick, \textit{Phys. Rev. Lett.} {\bf 74}, 2142 (1995).
\bibitem{Chuang} J. Chuang, Y. Kantor and M. Kardar, \textit{Phys. Rev. E} {\bf 65}, 011802 (2002).
\bibitem{Kantor} Y. Kantor and M. Kardar,  \textit{Phys. Rev. E} {\bf 69}, 021806 (2004).
\bibitem{Milchev} A. Milchev, K. Binder, and A. Bhattacharya, \textit{J. Chem. Phys.} {\bf 121}, 6042 (2004).
\bibitem{Luo1} K. F. Luo, T. Ala-Nissila, and S. C. Ying, \textit{J. Chem. Phys.} {\bf 124}, 034714 (2006).
\bibitem{Luo2} K. F. Luo, I. Huopaniemi, T. Ala-Nissila, and S. C. Ying, \textit{J. Chem. Phys.} {\bf 124}, 114704 (2006).
\bibitem{Chern} S.-S. Chern, A. E. Cardenas, and R. D. Coalson, \textit{J. Chem. Phys.} {\bf 115}, 7772 (2001).
\bibitem{Loebl} H. C. Loebl, R. Randel, S. P. Goodwin, and C. C. Matthai, \textit{Phys. Rev. E} {\bf 67}, 041913 (2003).
\bibitem{Randel} R. Randel, H. C. Loebl, and C. C. Matthai, \textit{Macromol. Theory Simul.} {\bf 13}, 387 (2004).
\bibitem{Lansac} Y. Lansac, P. K. Maiti, and M. A. Glaser, \textit{Polymer} {\bf 45}, 3099 (2004).
\bibitem{Kong} C. Y. Kong and M. Muthukumar, \textit{Electrophoresis} {\bf 23}, 2697 (2002).
\bibitem{Farkas} Z. Farkas, I. Derenyi, and T. Vicsek, \textit{J. Phys.: Condens. Matter} {\bf 15}, S1767 (2003).
\bibitem{Tian} P. Tian and G. D. Smith, \textit{J. Chem. Phys.}  {\bf 119}, 11475 (2003).
\bibitem{Zandi} R. Zandi, D. Reguera, J. Rudnick, and W. M. Gelbart, \textit{Proc. Natl. Acad. Sci. 
					U.S.A.} {\bf 100}, 8649 (2003).
\bibitem{Guo} L. Guo and E. Luijten, 
     \textit{Computer Simulation Studies in Condensed-Matter XVIII}
	 Edit by D. P. Landau, S. P. Lewis, and H.-B. Schuttler (Springer, Heidelberg, 2006).
\bibitem{de Gennes} P. G. de Gennes, 
			\textit{Scaling Concepts in Polymer Physics} (Cornell University Press, Ithaca, NY, 1979).
\bibitem{Doi} M. Doi, and S. F. Edwards, \textit{The Theory of Polymer Dynamics} (Clarendon, Oxford, 1986).
\bibitem{Ermak} D. L. Ermak and H. Buckholz, \textit{Journal of Computational Physics} {\bf 35}, 169-182 (1980).
\bibitem{Allen} M.P. Allen, D.J. Tildesley, {\it Computer Simulation of Liquids} (Oxford University Press, 1987).
\bibitem{Footnote} For strong forces, the translocation time as a function of electric field strength saturates.
\bibitem{Bhatt} A. Bhattacharay, private communication.
\end{thebibliography}
\end{document}